\documentclass[aps,prl,twocolumn,superscriptaddress,nofootinbib]{revtex4-2}
\usepackage{graphicx}
\usepackage{amsmath,amssymb}
\usepackage{xcolor}
\usepackage{hyperref}
\usepackage{cancel}
\hypersetup{colorlinks=true,linkcolor=blue!60!black,citecolor=blue!60!black,urlcolor=blue!60!black}

\newcommand{\eq}[1]{Eq.~(\ref{#1})}
\newcommand{\fig}[1]{Fig.~\ref{#1}}
\newcommand{\Sodd}{S_{\rm odd}}
\newcommand{\SOfour}{S_{{\rm O}(4)}}

\begin{document}

\title{The O(4)-breaking bubble}

\author{Guy Avraham}
\email{guy.avraham@weizmann.ac.il}
\affiliation{Department of Particle Physics and Astrophysics, Weizmann Institute of Science, Rehovot 7610001, Israel}
\author{Kfir Blum}
\email{kfir.blum@weizmann.ac.il}
\affiliation{Department of Particle Physics and Astrophysics, Weizmann Institute of Science, Rehovot 7610001, Israel}
\author{Omri Rosner}
\email{omri.rosner@weizmann.ac.il}
\affiliation{Department of Particle Physics and Astrophysics, Weizmann Institute of Science, Rehovot 7610001, Israel}
\author{Isaac G. Smith}
\email{isaacsmi@weizmann.ac.il}
\affiliation{Department of Particle Physics and Astrophysics, Weizmann Institute of Science, Rehovot 7610001, Israel}

\date{\today}

\begin{abstract}
False vacuum decay in scalar field theory is thought to be dominated by Coleman's O(4)-symmetric bounce, the minimum action nontrivial solution of the imaginary-time equations of motion. Beyond the bounce, non-constructive existence proofs of O(4)-breaking solutions are available in the mathematics literature, but the solutions themselves, and their physics, have remained unknown. Considering the simple, bounded-below, scalar field potential $V(\phi)=\frac{m^2}{2}\phi^2-\frac{\lambda}{4}\phi^4+\frac{g}{6}\phi^6$, we construct a nonradial solution explicitly: two bubble tubes of opposite sign wrapping orthogonal rings, invariant under ${\rm O}(2)\times{\rm O}(2)$ rotations combined with a $Z_2$ symmetry that exchanges the rings. The solution admits valid Cauchy data for real-time evolution from a $t=0$ slice, and supports an odd $>1$ number of unstable deformation modes. Having found this first solution we proceed to discover additional examples, in 3d as well as other 4d theories including a $Z_2$-violating potential and scalar QED for which there were no previously known existence proofs or O(4)-dominance theorems. 
\end{abstract}

\maketitle

\textit{Introduction.---}The semiclassical theory of false vacuum decay rests on a Euclidean saddle point: the bounce, a finite-action solution of the Euclidean (imaginary-time) equations of motion that interpolates through the potential barrier~\cite{Coleman:1977py,Callan:1977pt}. For a single scalar field in flat four-dimensional spacetime, Coleman, Glaser, and Martin (CGM) proved that the solution of minimum action is O(4) symmetric~\cite{Coleman:1977th}, and their proof was extended to multiple fields in~\cite{Blum:2016ipp}. The CGM theorem fixes the leading semiclassical decay rate, but it is a statement about the minimizer, not a uniqueness theorem: it leaves open whether the theory admits {\it any} other regular finite-action solutions of lower symmetry. Known examples of reduced-symmetry bounces rely on an external agent that breaks O(4) by hand---background fields~\cite{Hassan:2024nbl,Hassan:2024zkd}, finite temperature~\cite{Shoji:2025nvj}, a de Sitter horizon~\cite{Masoumi:2012rx}, an already-existing bubble wall~\cite{Czech:2011aa}, or singular field configurations sustained by tuned potentials~\cite{Sasaki:2024vul,Sasaki:2025nzy}. 
In the mathematics literature, by contrast, existence is settled under reasonably broad admissibility conditions. Bartsch and Willem~\cite{Bartsch:1993} proved that the Euclidean scalar field equation admits finite-action sign-changing solutions invariant under ${\rm O}(2)\times{\rm O}(2)$ and odd under the exchange of two orthogonal planes, and Mederski~\cite{Mederski:2017} and Jeanjean and Lu~\cite{Jeanjean:2018} extended the existence to the optimal (Berestycki--Lions) class of nonlinearities. These arguments permitted potentials growing no faster than quartically. However, a steeper, confining potential---such as the one we study below---is nonetheless brought into this class by truncation~\cite{Berestycki:1983}, a modification that leaves bounded finite-action solutions unchanged. These proofs were non-constructive: they characterized a minimizer within the symmetry class and bounded its action from below, but the solution itself---its profile, fluctuation spectrum, and physical role---was not exhibited. Constructive and spectral results do exist for neighboring classes: necklaces of alternating-sign bumps with dihedral symmetry~\cite{Musso:2012}, which are explicitly not of the present type, and, for the critical conformal equation, sign-changing solutions concentrating on Hopf links and Clifford tori~\cite{delPino:2013}. The Bartsch--Willem class itself has, to our knowledge, never been exhibited concretely. 

In this Letter we construct a solution, and read off its physics. After doing so first for the Bartsch--Willem class, we extend our method and discover other solutions that go beyond currently known existence proofs: ${\rm O}(2)\times{\rm O}(2)$ without parity symmetry; as well as solutions in scalar quantum electrodynamics (QED), where no CGM-like O(4)-dominance theorem exists. 

For concreteness, we start with the theory
\begin{equation}
V(\phi)=\frac{m^2}{2}\phi^2-\frac{\lambda}{4}\phi^4+\frac{g}{6}\phi^6,
\label{eq:V}
\end{equation}
with $m^2,\lambda,g>0$, bounded from below. Rescaling $\phi\to(m/\sqrt{\lambda})\,\phi$ and $x\to x/m$ brings the Euclidean action to
\begin{equation}
S=\frac{1}{\lambda}\int d^4x\left[\frac{1}{2}(\partial\phi)^2+\frac{1}{2}\phi^2-\frac{1}{4}\phi^4+\frac{\bar g}{6}\phi^6\right],
\label{eq:Sresc}
\end{equation}
with $\bar g=g m^2/\lambda^2$. At the leading semiclassical order, the theory depends on the single dimensionless combination $\bar g$, and the overall $1/\lambda$ cancels in any ratio of actions, in particular in $\Sodd/\SOfour$ below. We work with the rescaled form from here on and write $g$ for $\bar g$. The false vacuum sits at $\phi=0$; for $0<g<3/16$ a deeper vacuum exists at $\phi\neq0$, becoming degenerate with the origin as $g\to3/16$ (the thin-wall limit). 

\begin{figure}[t]
\includegraphics[width=0.95\columnwidth]{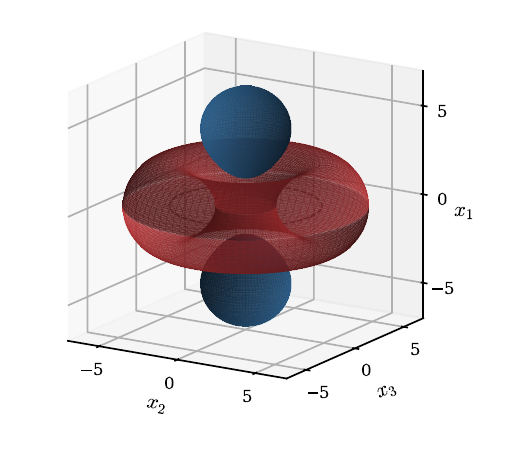}
\caption{Constant-field surfaces $|\phi|=1$ of the nonradial solution at $g=0.1$, on the slice $x_0=0$. The positive ring appears as a torus in the $(x_2,x_3)$ plane (core dashed); the negative ring, which lies in the $(x_0,x_1)$ plane, pierces the slice in two anti-sign bubbles on the torus axis at $x_1=\pm R_*$, with $R_*\simeq 4.15$. (The solution on the reduced $(r,s)$ half-plane is shown in the Supplemental Material (SM), Fig.~\ref{fig:solution}, left panel.) Because the solution is even in $x_0$, the $x_0=0$ slice carries valid Cauchy data for real-time development.}
\label{fig:slice}
\end{figure}

\textit{The solution.---}Group the four Euclidean coordinates into two planes and define $r=\sqrt{x_0^2+x_1^2}$, $s=\sqrt{x_2^2+x_3^2}$. We impose the ansatz
\begin{equation}
\phi=\phi(r,s),\qquad \phi(r,s)=-\phi(s,r),
\label{eq:ansatz}
\end{equation}
i.e.\ invariance under ${\rm O}(2)\times{\rm O}(2)$ rotations of the two planes, together with a parity $P$ that exchanges the planes and simultaneously flips $\phi\to-\phi$ (a symmetry of~\eq{eq:V}, which is even in $\phi$). This is precisely the symmetry class of Refs.~\cite{Bartsch:1993,Mederski:2017,Jeanjean:2018} (the space $X_\tau\cap H^1_{\mathcal{O}_2}$ in the notation of~\cite{Mederski:2017,Jeanjean:2018}, with $N=4$, $M=2$), which inspired the ansatz; a finite-action solution minimizing the action among all nontrivial solutions in this class is guaranteed to exist for our potential~\cite{Mederski:2017,Jeanjean:2018}. 
We solve the Euclidean equation $\nabla^2\phi=V'(\phi)$ on the $(r,s)$ half-plane with a globalized Newton method starting from a seed function in the symmetry class. The solution converges on multiple grids, to machine precision (see SM for details). 

\fig{fig:slice} shows the constant-field surfaces of the solution on the slice $x_0=0$ for $g=0.1$. The structure is transparent: a lump of positive field centered on the $s$ axis at $(r,s)=(0,R_*)$, and its $P$ mirror image, with negative sign, at $(R_*,0)$. Restoring the angles, the positive lump is a ``bubble tube'' whose core traces a ring of radius $R_*$ in the $(x_2,x_3)$ plane, while the negative tube wraps the orthogonal ring in the $(x_0,x_1)$ plane. The rings are mutually orthogonal in $\mathbb{R}^4$.  

The solution is even in $x_0$, so this slice carries time-symmetric Cauchy data $(\phi,\,\dot\phi=0)$, and conservation of the Euclidean-time energy along $x_0$, together with $\phi\to0$ as $x_0\to-\infty$, fixes the Lorentzian energy to zero. As for Coleman's bounce, the configuration would materialize at rest. Its Euclidean development and real-time evolution are computed in the SM.

\begin{figure}[t]
\includegraphics[width=0.95\columnwidth]{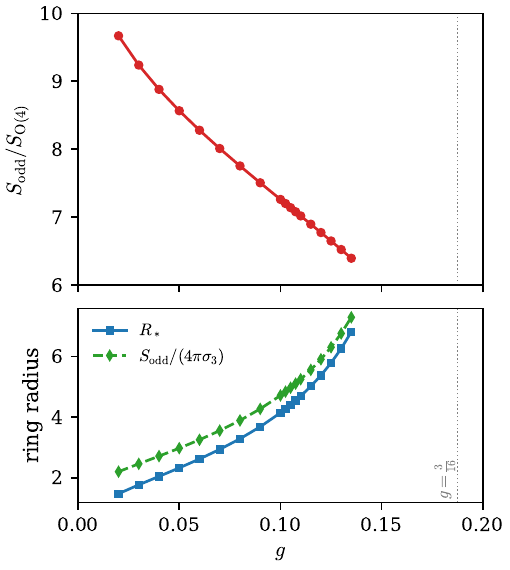}
\caption{Top: action of the nonradial solution in units of the O(4) bounce action, across the coupling range of the model. The ratio stays well above unity (the CGM bound), and indeed above the stronger bound $\Sodd\geq2\,\SOfour$ obeyed by every solution in the class. Bottom: ring radius $R_*$ (squares), compared with the effective radius $\Sodd/(4\pi\sigma_3)$ inferred from the thin-tube model~\eq{eq:tube} (diamonds). Numerical continuation fails for $g\gtrsim0.14$, where the thin-wall limit makes the Newton basin narrow; the trend of $R_*$ is consistent with the solution growing large and soft as $g\to3/16$ (dotted line).}
\label{fig:action}
\end{figure}

\textit{Action.---}At $g=0.1$ we find, after Richardson extrapolation in the grid spacing,
\begin{equation}
\Sodd=2849.8,\qquad
\frac{\Sodd}{\SOfour}=7.260,
\end{equation}
with $\SOfour=392.517$ the action of the O(4) bounce computed in the same units. The virial identity $S=K/2$, with $K$ the gradient term, is satisfied to six digits by the bounce and at the level of the $O(h^2)$ discretization error by the grid solution. We track the solution across $0.02\le g\le0.135$; the action ratio, shown in \fig{fig:action}, decreases monotonically from $9.7$ to $6.4$. The CGM~\cite{Coleman:1977th} bound, $\Sodd/\SOfour\geq1$, holds with room to spare, and so does the stronger variational bound $\Sodd\geq2\,\SOfour$, which every nontrivial solution of the symmetry class~\eq{eq:ansatz} must obey~\cite{Mederski:2017,Jeanjean:2018}. The latter bound has important consequences for physics: since nucleation rates scale as $e^{-S}$, any nonradial bubble in the symmetry class---including the solution constructed here---is less likely to nucleate than a \emph{pair} of independent Coleman bubbles. 

The tube cross section is, to good accuracy, the O(3)-symmetric bounce of the same potential: at $g=0.1$ the core value is $\phi=2.950$ against $\phi_3(0)=2.906$ for the three-dimensional bounce, whose action sets a tension $\sigma_3=48.05$. A thin-tube estimate of the action, two rings of length $2\pi R_*$ each,
\begin{equation}
\Sodd\approx 2\times 2\pi R_*\,\sigma_3,
\label{eq:tube}
\end{equation}
reproduces the true action to $O(10\%)$ at $g=0.1$. 
The ring radius grows from $R_*=1.5$ to $6.8$ across the same range, tracking the thin-tube estimate. Toward $g\to0$ the rings shrink and approach each other, and the tube picture degrades; toward the thin-wall limit the numerics lose the solution, and we do not know whether the branch persists all the way to $g=3/16$.

\textit{Fluctuation spectrum.---}The physical role of a saddle is fixed by the spectrum of its fluctuation operator, $\mathcal{H}=-\nabla^2+V''(\phi)$. Decomposing fluctuations as $\delta\phi=\psi(r,s)\,e^{i(m\theta_1+n\theta_2)}$, with $\theta_{1,2}$ the angles of the two planes, $\mathcal{H}$ block-diagonalizes into sectors labeled by $(m,n)$, with sectors $(m,n)$ and $(n,m)$ degenerate by $P$. Table~\ref{tab:spectrum} summarizes the low-lying spectrum at $g=0.1$.

\begin{table}[b]
\caption{Low-lying fluctuation spectrum at $g=0.1$. Degeneracies include the angular multiplicity $(2-\delta_{m0})(2-\delta_{n0})$ and the $(m,n)\leftrightarrow(n,m)$ exchange. At this coupling there are 15 negative and 8 zero modes. All $\omega^2$ carry an $O(h^2)$ discretization uncertainty below $10^{-5}$, estimated by Richardson extrapolation to vanishing grid spacing.}
\label{tab:spectrum}
\begin{ruledtabular}
\begin{tabular}{lccl}
sector & $\omega^2$ & degeneracy & interpretation \\
\hline
$(0,0)$ & $-0.492$ & 1 & breathing, $P$-even \\
$(0,0)$ & $-0.315$ & 1 & breathing, $P$-odd (dilation) \\
$(0,0)$ & $-0.090$ & 1 & ring--ring see-saw \\
$(0,1)$ & $-0.349$ & 4 & tube bending, $n=1$ \\
$(0,2)$ & $-0.200$ & 4 & tube bending, $n=2$ \\
$(0,3)$ & $-0.0078$ & 4 & tube bending, $n=3$ \\
$(0,1)$ & $0$ & 4 & broken translations \\
$(1,1)$ & $0$ & 4 & broken O(4) rotations \\
$(0,4)$ & $+0.233$ & 4 & first stable bending mode \\
\end{tabular}
\end{ruledtabular}
\end{table}

The zero modes count the broken symmetries and close the bookkeeping: four translations, plus four rotations, because the solution breaks the six generators of O(4) down to the two of ${\rm O}(2)\times{\rm O}(2)$. We verified that the $(1,1)$ quadruplet of zero modes are proportional to the combination $r\partial_s\phi-s\partial_r\phi$ to numerical precision. 

The negative modes have an equally clean origin. A straight, isolated bubble tube has a single localized negative mode, its cross-sectional collapse/expansion, with $\omega_0^2=-0.561$ at $g=0.1$. Bending the tube into a ring of radius $R_*$ and Fourier decomposing along it gives the dispersion estimate
\begin{equation}
\omega_n^2\approx\omega_0^2+\frac{n^2}{R_*^2},
\label{eq:dispersion}
\end{equation}
which predicts instability for $n\le3$ and stability for $n\ge4$ at $g=0.1$---exactly as found in Table~\ref{tab:spectrum}, where the $n=3$ mode sits barely below zero and the $n=4$ mode safely above. (The estimate~\eq{eq:dispersion} reproduces the eigenvalues themselves only to within a factor of two for the lower modes; its robust output is the location of the zero crossing.) The two $n=0$ breathing modes of the individual rings hybridize into the $P$-even/$P$-odd pair at the bottom of the $(0,0)$ sector; the $P$-odd combination overlaps at the $93\%$ level with the dilation direction $x\cdot\nabla\phi$, i.e.\ it is the ``would-be bounce'' negative mode of this saddle. The soft third $(0,0)$ mode is a see-saw that grows one ring at the expense of the other.

Following the spectrum along the branch (SM \fig{fig:modes}) turns the mode count into a function of $g$. The dispersion~\eq{eq:dispersion} makes the trend intuitive: as $g$ grows, $R_*$ grows and $|\omega_0^2|$ shrinks, so bending modes are stabilized one quadruplet at a time. The $n=4$ quadruplet crosses zero at $g\simeq0.034$ and the $n=3$ quadruplet at $g\simeq0.104$; the count of negative modes steps through $19\to15\to11$, while the 8 zero modes persist unchanged, as they must. We checked that the marginal eigenvalues are stable against box size and grid spacing at the crossings (SM). 

\textit{What this saddle is, and is not.---}The O(4)-breaking solution does not contribute to the imaginary part of the false-vacuum energy at leading semiclassical order, as its action is well above the O(4) bounce's action~\cite{Callan:1977pt} and, moreover, it has multiple fluctuation eigenmodes~\cite{Coleman:1987rm}. It does populate the subleading structure of the path integral, related to the large-order behavior of the theory~\cite{Basar:2013eka}.

Whether the saddle contributes to vacuum decay at \emph{any} order depends, in Picard--Lefschetz terms~\cite{Witten:2010cx,Tanizaki:2014xba,Behtash:2015loa}, on whether its upward-flow cycle intersects the integration cycle of the false-vacuum path integral. Two facts argue that it does. First, the negative-mode count $k$ is odd everywhere on the branch (19, 15, or 11---structurally: three modes from the $(0,0)$ sector plus quadruplets), so the fluctuation factor $i^k$ is imaginary and could feed into ${\rm Im}\,E$. Second, the gradient flow through real-field configuration space along the $P$-odd dilation mode connects the false vacuum on one side to unbounded true-vacuum growth on the other (SM): the saddle behaves like a genuine transition state, with the same flow topology that underlies the bounce's contribution. The other two $(0,0)$ negative modes connect escape to escape; we did not flow the remaining 12 negative modes, as this would require three-dimensional flows.

\textit{Gravitational waves.---}Coleman's bubble is spherical and radiates gravitationally only through quantum fluctuations~\cite{Blum:2024hcs}. The nonradial bubble radiates already at leading classical order. We can make a rough estimate of the emission using the quadrupole formula. The wall tension is $\sigma\sim m^3/\lambda$, arising from the field swinging by $\Delta\phi\sim m/\sqrt{\lambda}$ across a thickness $\sim1/m$; the nucleated configuration is an oblate ring of this wall, of radius $R_*$ and energy $M\sim\sigma R_*^2$, whose mass quadrupole $Q\sim MR_*^2$ changes over the light-crossing time $t_*\sim R_*$ as the bubble rounds out. Then
\begin{equation}
E_{\rm GW}\sim G\,\frac{M^2}{R_*}\sim G\,\sigma^2 R_*^3\sim\left(\frac{m}{M_{\rm Pl}}\right)^2\lambda^{-2}\,m,
\label{eq:gwest}
\end{equation}
where $G=M_{\rm Pl}^{-2}$. The radiated energy is a fraction $E_{\rm GW}/M\sim G\sigma R_*\sim(m/M_{\rm Pl})^2/\lambda$ of the bubble's energy budget. Against the fluctuation-induced emission of a \emph{radial} bubble~\cite{Blum:2024hcs}, which scales as a single power of $\sigma$ where ours carries $\sigma^2$, the classical signal is parametrically enhanced by $1/\lambda$. A direct calculation of $E_{\rm GW}$ using the numerical real-time evaluation is a feasible task, in principle, but is beyond the scope of this work. We expect that, weighted by the nucleation probability $e^{-(\Sodd-\SOfour)}$, the nonradial population's emission should be negligible in any stochastic background.

\begin{figure}[t]
\includegraphics[width=0.95\columnwidth]{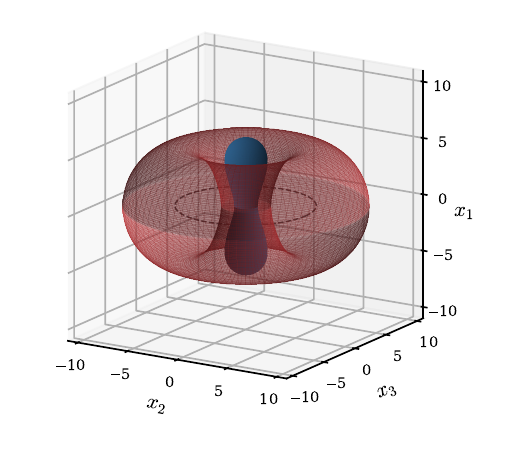}
\caption{Constant-field surfaces $|\phi|=1$ on the slice $x_0=0$ of the nonradial, parity-violating solution obtained for the potential of \eq{eq:kappa_potential} at $\kappa=g=0.1$.}
\label{fig:slicek}
\end{figure}

\textit{Other solutions: (i) parity-violating potential.---}The choice of \eq{eq:V} is only representative: the same machinery allows us to construct nonradial solutions in other potentials, including solutions for which, to our knowledge, there have not been earlier proofs of existence. Consider removing the $Z_2$ symmetry of~\eq{eq:V}, using the potential:
\begin{equation}
V_\kappa(\phi)=\frac{1}{2}\phi^2-\frac{1}{4}\phi^4+\frac{g}{6}\phi^6
+\kappa\phi^3.
\label{eq:kappa_potential}
\end{equation}
For $\kappa=g=0.1$, starting with the $\kappa=0$ nonradial solution as seed, we find a new solution. The $x_0=0$ slice of this solution is shown in Fig.~\ref{fig:slicek} (also SM Fig.~\ref{fig:solution}, right panel). The main features of this new saddle are consistent with a deformation of the $Z_2$-symmetric case: at the chosen values of $g$ and $\kappa$ it has 13 negative modes, and action larger than that of both O(4)-symmetric saddles that the theory admits. 

\begin{figure}[t]
\includegraphics[width=0.95\columnwidth]{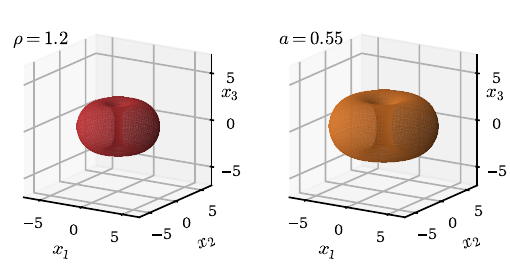}
\caption{Constant-field surfaces on the slice $x_0=0$ of the nonradial scalar QED $(1,0)$ saddle for $g=0.1$ and $e=1$.}
\label{fig:fig5_sqed}
\end{figure}

\begin{figure*}[t]
\includegraphics[width=1.5\columnwidth]{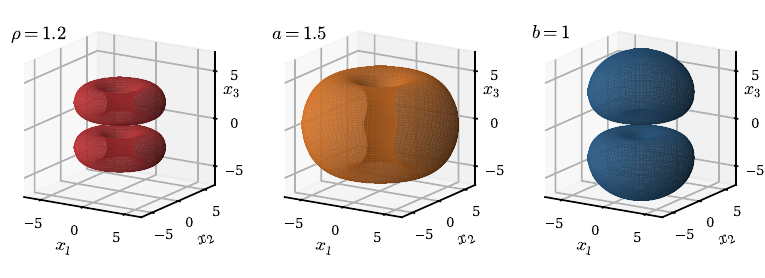}
\caption{Constant-field surfaces on the slice $x_0=0$ of the $(m,n)=(3,2)$ nonradial scalar QED saddle for $g=0.1$ and $e=1$.}
\label{fig:fig6_sqed32}
\end{figure*}

\textit{Other solutions: (ii) scalar QED.---}Here,
\begin{equation}
S=\int d^4x\left[
\frac{1}{4e^2}F_{\mu\nu}F_{\mu\nu}
+|D_\mu\Phi|^2+V(\rho)\right],\;\;
\Phi=\frac{\rho}{\sqrt{2}}e^{i\theta},
\label{eq:sqed_action}
\end{equation}
with $V(\rho)=\frac{1}{2}\rho^2-{1\over4}\rho^4+{g\over6}\rho^6$ and $D_\mu=\partial_\mu-iA_\mu$. 
Defining $(x_0,x_1)=r(\cos\alpha,\sin\alpha)$ and $(x_2,x_3)=s(\cos\beta,\sin\beta)$, 
we find O(4)-breaking saddles starting from the ansatz\footnote{Ref.~\cite{Liu:2025} used a similar construction to find infinite-action solutions of the magnetic Ginzburg-Landau equations in $4d$.}:
\begin{equation}
\Phi_B=\frac{\rho(r,s)}{\sqrt{2}}\,e^{i(m\alpha+n\beta)},
\;\;
A_B=a(r,s)d\alpha+b(r,s)d\beta,
\label{eq:sqed_o2o2_ansatz}
\end{equation}
with integer $m,n$. The lowest-action branch we found is $(m,n)=(1,0)$ or $(0,1)$, demonstrated in Fig.~\ref{fig:fig5_sqed} for $g=0.1$ and $e=1$. Here we can reduce \eq{eq:sqed_o2o2_ansatz} to $\Phi_B={\rho(r,s)\over\sqrt2}\,e^{i\alpha}$, 
$A_B=a(r,s)d\alpha$. Interestingly, for scalar QED there is no O(4)-dominance theorem. The solution's action decreases with increasing gauge coupling $e$: for $e=1$ we find $S_{(1,0)}(e=1)/S_{\rm O(4)}\approx2.120$, while for $e=4$ we find $S_{(1,0)}(e=4)/S_{\rm O(4)}\approx1.708$. Thus, the $\Sodd\geq2\,\SOfour$ bound of the pure scalar theory~\cite{Mederski:2017,Jeanjean:2018} indeed does not apply. Increasing $e$ causes the radius of the null tube threading $\rho(r,s)$ and $a(r,s)$ (see \fig{fig:fig5_sqed}) to shrink, which requires numerical care and we did not pursue it beyond $e=4$.   
As another example, Fig.~\ref{fig:fig6_sqed32} shows the $x_0=0$ slice of an $(m,n)=(3,2)$ saddle with $S_{(3,2)}(e=1)/S_{\rm O(4)}\approx10.74$. For the scalar QED solutions we did not attempt to count the number of negative modes $N_{-}$, but for all examples we were able to establish $N_{-}\geq3$. 

We note an interesting difference between the pure scalar and the scalar QED scenarios. The existence of a pure scalar finite-action nonradial saddle seems to require at least three potential local minima: the false vacuum and two other minima between which the solution interpolates; Eqs.~(\ref{eq:V}) and~(\ref{eq:kappa_potential}) were minimal examples in this sense. The reason for this is the Gidas, Ni \& Nirenberg theorem~\cite{Gidas:1979} which states that decaying solutions that do not change sign must be radial. Scalar QED does not adhere to this rule: the phase of the scalar field is gauge-dependent. Indeed, although we do not expand on this here, we were able to find a nonradial solution of the $(1,0)$ type in scalar QED with the double-well potential $V(\rho)=\frac{1}{2}\rho^2-\frac{1}{3}\rho^3+\frac{\lambda}{4}\rho^4$ for $\lambda=0.15$ and $e=1$. 

\begin{figure}[t]
\includegraphics[width=0.95\columnwidth]{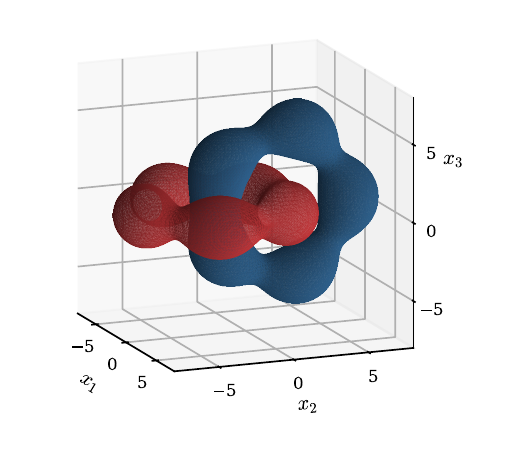}
\caption{Constant-field $|\phi|=1$ surfaces of the O(3)-breaking saddle in $d=3$ Euclidean dimensions.}
\label{fig:fig4_hiT}
\end{figure}

\textit{Other solutions: (iii) 3d.---}It is of interest to explore nonradial saddles in different numbers of dimensions. We demonstrate an O(3)-breaking saddle in $d=3$ with the potential \eq{eq:V}; this could refer to the tunneling state in $d=2+1$ (where, incidentally, the $\phi^6$ theory is renormalizable), or to the high-temperature dimensionally-reduced theory starting from $d=3+1$. With intuition gained in the $4d$ analysis, we construct a seed configuration made of positive and negative linked tubes with tube cross section set by the O(2)-symmetric bounce of the same potential. For $g=0.1$ this seed converges into the saddle shown in \fig{fig:fig4_hiT}. Interestingly, the tubes partially fragment into a linked five-bead pattern. The action (energy in the high-T interpretation) of this saddle is $S_{\cancel{\rm O(3)}}/S_{\rm O(3)}\approx10.238$ and it has $N_{-}>1$.

\textit{Outlook.---}The first solution we presented gives concrete form to the nonradial solutions whose existence was established variationally in~\cite{Bartsch:1993,Mederski:2017,Jeanjean:2018}, demonstrating that flat-space, zero-temperature scalar field theory admits regular Euclidean saddles beyond the O(4) bounce, with action an order-one multiple of the bounce action. Using the same machinery we also presented -- in lesser detail -- new classes of solutions that go beyond previous existence proofs. This is likely the tip of the iceberg of reduced-symmetry saddles, the role of which has been speculated for five decades~\cite{Coleman:1977py,Coleman:1977th,Blum:2016ipp}.
It would be interesting to pursue generalizations to different numbers of dimensions, compact or periodic boundary conditions, other potentials, and multi-field models, where the landscape of saddles is essentially unexplored. It would also be interesting to extend the investigation of scalar gauge theories, and especially to understand if O(4)-dominance theorems~\cite{Coleman:1977th,Blum:2016ipp,Mederski:2017,Jeanjean:2018} apply in that arena. 
Our methods should also be useful for parallel investigations in condensed matter physics; see, e.g.~\cite{Zhang:2025,Benfenati:2019}.

\textit{Acknowledgments.---}We thank Ofer Aharony and Ryosuke Sato for comments on the ms; Ryosuke, in particular, prompted us to experiment with additional classes of potentials. We acknowledge support by the UNDARK project (project number 101159929), and
by the Minerva foundation with funding from the Federal German Ministry for Education
and Research. 

\vspace{2mm}
Numerical scripts reproducing all results of this Letter are available from the authors.

\bibliography{ref}

\begin{thebibliography}{27}%
\makeatletter
\providecommand \@ifxundefined [1]{%
 \@ifx{#1\undefined}
}%
\providecommand \@ifnum [1]{%
 \ifnum #1\expandafter \@firstoftwo
 \else \expandafter \@secondoftwo
 \fi
}%
\providecommand \@ifx [1]{%
 \ifx #1\expandafter \@firstoftwo
 \else \expandafter \@secondoftwo
 \fi
}%
\providecommand \natexlab [1]{#1}%
\providecommand \enquote  [1]{``#1''}%
\providecommand \bibnamefont  [1]{#1}%
\providecommand \bibfnamefont [1]{#1}%
\providecommand \citenamefont [1]{#1}%
\providecommand \href@noop [0]{\@secondoftwo}%
\providecommand \href [0]{\begingroup \@sanitize@url \@href}%
\providecommand \@href[1]{\@@startlink{#1}\@@href}%
\providecommand \@@href[1]{\endgroup#1\@@endlink}%
\providecommand \@sanitize@url [0]{\catcode `\\12\catcode `\$12\catcode
  `\&12\catcode `\#12\catcode `\^12\catcode `\_12\catcode `\%12\relax}%
\providecommand \@@startlink[1]{}%
\providecommand \@@endlink[0]{}%
\providecommand \url  [0]{\begingroup\@sanitize@url \@url }%
\providecommand \@url [1]{\endgroup\@href {#1}{\urlprefix }}%
\providecommand \urlprefix  [0]{URL }%
\providecommand \Eprint [0]{\href }%
\providecommand \doibase [0]{https://doi.org/}%
\providecommand \selectlanguage [0]{\@gobble}%
\providecommand \bibinfo  [0]{\@secondoftwo}%
\providecommand \bibfield  [0]{\@secondoftwo}%
\providecommand \translation [1]{[#1]}%
\providecommand \BibitemOpen [0]{}%
\providecommand \bibitemStop [0]{}%
\providecommand \bibitemNoStop [0]{.\EOS\space}%
\providecommand \EOS [0]{\spacefactor3000\relax}%
\providecommand \BibitemShut  [1]{\csname bibitem#1\endcsname}%
\let\auto@bib@innerbib\@empty
\bibitem [{\citenamefont {Coleman}(1977)}]{Coleman:1977py}%
  \BibitemOpen
  \bibfield  {author} {\bibinfo {author} {\bibfnamefont {S.~R.}\ \bibnamefont
  {Coleman}},\ }\bibfield  {title} {\bibinfo {title} {{The Fate of the False
  Vacuum. 1. Semiclassical Theory}},\ }\href
  {https://doi.org/10.1103/PhysRevD.15.2929} {\bibfield  {journal} {\bibinfo
  {journal} {Phys. Rev. D}\ }\textbf {\bibinfo {volume} {15}},\ \bibinfo
  {pages} {2929} (\bibinfo {year} {1977})},\ \bibinfo {note} {[Erratum:
  Phys.Rev.D 16, 1248 (1977)]}\BibitemShut {NoStop}%
\bibitem [{\citenamefont {Callan}\ and\ \citenamefont
  {Coleman}(1977)}]{Callan:1977pt}%
  \BibitemOpen
  \bibfield  {author} {\bibinfo {author} {\bibfnamefont {C.~G.}\ \bibnamefont
  {Callan}, \bibfnamefont {Jr.}}\ and\ \bibinfo {author} {\bibfnamefont
  {S.~R.}\ \bibnamefont {Coleman}},\ }\bibfield  {title} {\bibinfo {title}
  {{The Fate of the False Vacuum. 2. First Quantum Corrections}},\ }\href
  {https://doi.org/10.1103/PhysRevD.16.1762} {\bibfield  {journal} {\bibinfo
  {journal} {Phys. Rev. D}\ }\textbf {\bibinfo {volume} {16}},\ \bibinfo
  {pages} {1762} (\bibinfo {year} {1977})}\BibitemShut {NoStop}%
\bibitem [{\citenamefont {Coleman}\ \emph {et~al.}(1978)\citenamefont
  {Coleman}, \citenamefont {Glaser},\ and\ \citenamefont
  {Martin}}]{Coleman:1977th}%
  \BibitemOpen
  \bibfield  {author} {\bibinfo {author} {\bibfnamefont {S.~R.}\ \bibnamefont
  {Coleman}}, \bibinfo {author} {\bibfnamefont {V.}~\bibnamefont {Glaser}},\
  and\ \bibinfo {author} {\bibfnamefont {A.}~\bibnamefont {Martin}},\
  }\bibfield  {title} {\bibinfo {title} {{Action Minima Among Solutions to a
  Class of Euclidean Scalar Field Equations}},\ }\href
  {https://doi.org/10.1007/BF01609421} {\bibfield  {journal} {\bibinfo
  {journal} {Commun. Math. Phys.}\ }\textbf {\bibinfo {volume} {58}},\ \bibinfo
  {pages} {211} (\bibinfo {year} {1978})}\BibitemShut {NoStop}%
\bibitem [{\citenamefont {Blum}\ \emph {et~al.}(2017)\citenamefont {Blum},
  \citenamefont {Honda}, \citenamefont {Sato}, \citenamefont {Takimoto},\ and\
  \citenamefont {Tobioka}}]{Blum:2016ipp}%
  \BibitemOpen
  \bibfield  {author} {\bibinfo {author} {\bibfnamefont {K.}~\bibnamefont
  {Blum}}, \bibinfo {author} {\bibfnamefont {M.}~\bibnamefont {Honda}},
  \bibinfo {author} {\bibfnamefont {R.}~\bibnamefont {Sato}}, \bibinfo {author}
  {\bibfnamefont {M.}~\bibnamefont {Takimoto}},\ and\ \bibinfo {author}
  {\bibfnamefont {K.}~\bibnamefont {Tobioka}},\ }\bibfield  {title} {\bibinfo
  {title} {{O($N$) Invariance of the Multi-Field Bounce}},\ }\href
  {https://doi.org/10.1007/JHEP05(2017)109} {\bibfield  {journal} {\bibinfo
  {journal} {JHEP}\ }\textbf {\bibinfo {volume} {05}},\ \bibinfo {pages}
  {109}},\ \bibinfo {note} {[Erratum: JHEP 06, 060 (2017)]},\ \Eprint
  {https://arxiv.org/abs/1611.04570} {arXiv:1611.04570 [hep-th]} \BibitemShut
  {NoStop}%
\bibitem [{\citenamefont {Hassan}\ \emph {et~al.}(2024)\citenamefont {Hassan},
  \citenamefont {March-Russell},\ and\ \citenamefont {Obied}}]{Hassan:2024nbl}%
  \BibitemOpen
  \bibfield  {author} {\bibinfo {author} {\bibfnamefont {S.}~\bibnamefont
  {Hassan}}, \bibinfo {author} {\bibfnamefont {J.}~\bibnamefont
  {March-Russell}},\ and\ \bibinfo {author} {\bibfnamefont {G.}~\bibnamefont
  {Obied}},\ }\bibfield  {title} {\bibinfo {title} {{Chern-Simons bubbles:
  Lopsided false vacuum decay in axion electrodynamics}},\ }\href@noop {} {\
  (\bibinfo {year} {2024})},\ \Eprint {https://arxiv.org/abs/2402.00119}
  {arXiv:2402.00119 [hep-th]} \BibitemShut {NoStop}%
\bibitem [{\citenamefont {Hassan}(2024)}]{Hassan:2024zkd}%
  \BibitemOpen
  \bibfield  {author} {\bibinfo {author} {\bibfnamefont {S.}~\bibnamefont
  {Hassan}},\ }\emph {\bibinfo {title} {{Topological bubble nucleation}}},\
  \href {https://doi.org/10.5287/ora-kkxyzk5kr} {Ph.D. thesis},\ \bibinfo
  {school} {Oxford University} (\bibinfo {year} {2024})\BibitemShut {NoStop}%
\bibitem [{\citenamefont {Shoji}\ and\ \citenamefont
  {Yamaguchi}(2025)}]{Shoji:2025nvj}%
  \BibitemOpen
  \bibfield  {author} {\bibinfo {author} {\bibfnamefont {Y.}~\bibnamefont
  {Shoji}}\ and\ \bibinfo {author} {\bibfnamefont {M.}~\bibnamefont
  {Yamaguchi}},\ }\bibfield  {title} {\bibinfo {title} {{Symmetry of Bounce
  Solutions at Finite Temperature}},\ }\href@noop {} {\  (\bibinfo {year}
  {2025})},\ \Eprint {https://arxiv.org/abs/2511.05950} {arXiv:2511.05950
  [hep-th]} \BibitemShut {NoStop}%
\bibitem [{\citenamefont {Masoumi}\ and\ \citenamefont
  {Weinberg}(2012)}]{Masoumi:2012rx}%
  \BibitemOpen
  \bibfield  {author} {\bibinfo {author} {\bibfnamefont {A.}~\bibnamefont
  {Masoumi}}\ and\ \bibinfo {author} {\bibfnamefont {E.~J.}\ \bibnamefont
  {Weinberg}},\ }\bibfield  {title} {\bibinfo {title} {{Bounces with O(3) x
  O(2) symmetry}},\ }\href {https://doi.org/10.1103/PhysRevD.86.104029}
  {\bibfield  {journal} {\bibinfo  {journal} {Phys. Rev. D}\ }\textbf {\bibinfo
  {volume} {86}},\ \bibinfo {pages} {104029} (\bibinfo {year} {2012})},\
  \Eprint {https://arxiv.org/abs/1207.3717} {arXiv:1207.3717 [hep-th]}
  \BibitemShut {NoStop}%
\bibitem [{\citenamefont {Czech}(2012)}]{Czech:2011aa}%
  \BibitemOpen
  \bibfield  {author} {\bibinfo {author} {\bibfnamefont {B.}~\bibnamefont
  {Czech}},\ }\bibfield  {title} {\bibinfo {title} {{Barnacles -- A Novel
  Channel for Vacuum Decay}},\ }\href
  {https://doi.org/10.1016/j.physletb.2012.06.018} {\bibfield  {journal}
  {\bibinfo  {journal} {Phys. Lett. B}\ }\textbf {\bibinfo {volume} {713}},\
  \bibinfo {pages} {331} (\bibinfo {year} {2012})},\ \Eprint
  {https://arxiv.org/abs/1112.1638} {arXiv:1112.1638 [hep-th]} \BibitemShut
  {NoStop}%
\bibitem [{\citenamefont {Sasaki}\ \emph
  {et~al.}(2025{\natexlab{a}})\citenamefont {Sasaki}, \citenamefont
  {Yingcharoenrat},\ and\ \citenamefont {Zhang}}]{Sasaki:2024vul}%
  \BibitemOpen
  \bibfield  {author} {\bibinfo {author} {\bibfnamefont {M.}~\bibnamefont
  {Sasaki}}, \bibinfo {author} {\bibfnamefont {V.}~\bibnamefont
  {Yingcharoenrat}},\ and\ \bibinfo {author} {\bibfnamefont {Y.-l.}\
  \bibnamefont {Zhang}},\ }\bibfield  {title} {\bibinfo {title} {{Beyond
  Coleman's Instantons}},\ }\href {https://doi.org/10.1088/1361-6382/ae2376}
  {\bibfield  {journal} {\bibinfo  {journal} {Class. Quant. Grav.}\ }\textbf
  {\bibinfo {volume} {42}},\ \bibinfo {pages} {235017} (\bibinfo {year}
  {2025}{\natexlab{a}})},\ \Eprint {https://arxiv.org/abs/2411.11322}
  {arXiv:2411.11322 [hep-th]} \BibitemShut {NoStop}%
\bibitem [{\citenamefont {Sasaki}\ \emph
  {et~al.}(2025{\natexlab{b}})\citenamefont {Sasaki}, \citenamefont
  {Yingcharoenrat},\ and\ \citenamefont {Zhang}}]{Sasaki:2025nzy}%
  \BibitemOpen
  \bibfield  {author} {\bibinfo {author} {\bibfnamefont {M.}~\bibnamefont
  {Sasaki}}, \bibinfo {author} {\bibfnamefont {V.}~\bibnamefont
  {Yingcharoenrat}},\ and\ \bibinfo {author} {\bibfnamefont {Y.-l.}\
  \bibnamefont {Zhang}},\ }\bibfield  {title} {\bibinfo {title} {{Singular
  instantons with finite action}},\ }\href
  {https://doi.org/10.1142/S0217751X2540010X} {\bibfield  {journal} {\bibinfo
  {journal} {Int. J. Mod. Phys. A}\ }\textbf {\bibinfo {volume} {40}},\
  \bibinfo {pages} {2540010} (\bibinfo {year} {2025}{\natexlab{b}})},\ \Eprint
  {https://arxiv.org/abs/2503.19616} {arXiv:2503.19616 [hep-th]} \BibitemShut
  {NoStop}%
\bibitem [{\citenamefont {Bartsch}\ and\ \citenamefont
  {Willem}(1993)}]{Bartsch:1993}%
  \BibitemOpen
  \bibfield  {author} {\bibinfo {author} {\bibfnamefont {T.}~\bibnamefont
  {Bartsch}}\ and\ \bibinfo {author} {\bibfnamefont {M.}~\bibnamefont
  {Willem}},\ }\bibfield  {title} {\bibinfo {title} {{Infinitely many nonradial
  solutions of a Euclidean scalar field equation}},\ }\href
  {https://doi.org/10.1006/jfan.1993.1133} {\bibfield  {journal} {\bibinfo
  {journal} {J. Funct. Anal.}\ }\textbf {\bibinfo {volume} {117}},\ \bibinfo
  {pages} {447} (\bibinfo {year} {1993})}\BibitemShut {NoStop}%
\bibitem [{\citenamefont {Mederski}(2020)}]{Mederski:2017}%
  \BibitemOpen
  \bibfield  {author} {\bibinfo {author} {\bibfnamefont {J.}~\bibnamefont
  {Mederski}},\ }\bibfield  {title} {\bibinfo {title} {{Nonradial solutions of
  nonlinear scalar field equations}},\ }\href
  {https://doi.org/10.1088/1361-6544/aba889} {\bibfield  {journal} {\bibinfo
  {journal} {Nonlinearity}\ }\textbf {\bibinfo {volume} {33}},\ \bibinfo
  {pages} {6349} (\bibinfo {year} {2020})},\ \Eprint
  {https://arxiv.org/abs/1711.05711} {arXiv:1711.05711 [math.AP]} \BibitemShut
  {NoStop}%
\bibitem [{\citenamefont {Jeanjean}\ and\ \citenamefont
  {Lu}(2020)}]{Jeanjean:2018}%
  \BibitemOpen
  \bibfield  {author} {\bibinfo {author} {\bibfnamefont {L.}~\bibnamefont
  {Jeanjean}}\ and\ \bibinfo {author} {\bibfnamefont {S.-S.}\ \bibnamefont
  {Lu}},\ }\bibfield  {title} {\bibinfo {title} {{Nonlinear scalar field
  equations with general nonlinearity}},\ }\href
  {https://doi.org/10.1016/j.na.2019.111604} {\bibfield  {journal} {\bibinfo
  {journal} {Nonlinear Anal.}\ }\textbf {\bibinfo {volume} {190}},\ \bibinfo
  {pages} {111604} (\bibinfo {year} {2020})},\ \Eprint
  {https://arxiv.org/abs/1807.07350} {arXiv:1807.07350 [math.AP]} \BibitemShut
  {NoStop}%
\bibitem [{\citenamefont {Berestycki}\ and\ \citenamefont
  {Lions}(1983)}]{Berestycki:1983}%
  \BibitemOpen
  \bibfield  {author} {\bibinfo {author} {\bibfnamefont {H.}~\bibnamefont
  {Berestycki}}\ and\ \bibinfo {author} {\bibfnamefont {P.-L.}\ \bibnamefont
  {Lions}},\ }\bibfield  {title} {\bibinfo {title} {{Nonlinear scalar field
  equations, I: Existence of a ground state}},\ }\href
  {https://doi.org/10.1007/BF00250555} {\bibfield  {journal} {\bibinfo
  {journal} {Arch. Ration. Mech. Anal.}\ }\textbf {\bibinfo {volume} {82}},\
  \bibinfo {pages} {313} (\bibinfo {year} {1983})}\BibitemShut {NoStop}%
\bibitem [{\citenamefont {Musso}\ \emph {et~al.}(2012)\citenamefont {Musso},
  \citenamefont {Pacard},\ and\ \citenamefont {Wei}}]{Musso:2012}%
  \BibitemOpen
  \bibfield  {author} {\bibinfo {author} {\bibfnamefont {M.}~\bibnamefont
  {Musso}}, \bibinfo {author} {\bibfnamefont {F.}~\bibnamefont {Pacard}},\ and\
  \bibinfo {author} {\bibfnamefont {J.}~\bibnamefont {Wei}},\ }\bibfield
  {title} {\bibinfo {title} {{Finite-energy sign-changing solutions with
  dihedral symmetry for the stationary nonlinear Schr{\"o}dinger equation}},\
  }\href {https://doi.org/10.4171/JEMS/351} {\bibfield  {journal} {\bibinfo
  {journal} {J. Eur. Math. Soc.}\ }\textbf {\bibinfo {volume} {14}},\ \bibinfo
  {pages} {1923} (\bibinfo {year} {2012})}\BibitemShut {NoStop}%
\bibitem [{\citenamefont {del Pino}\ \emph {et~al.}(2013)\citenamefont {del
  Pino}, \citenamefont {Musso}, \citenamefont {Pacard},\ and\ \citenamefont
  {Pistoia}}]{delPino:2013}%
  \BibitemOpen
  \bibfield  {author} {\bibinfo {author} {\bibfnamefont {M.}~\bibnamefont {del
  Pino}}, \bibinfo {author} {\bibfnamefont {M.}~\bibnamefont {Musso}}, \bibinfo
  {author} {\bibfnamefont {F.}~\bibnamefont {Pacard}},\ and\ \bibinfo {author}
  {\bibfnamefont {A.}~\bibnamefont {Pistoia}},\ }\bibfield  {title} {\bibinfo
  {title} {{Torus action on $S^n$ and sign changing solutions for conformally
  invariant equations}},\ }\href {https://doi.org/10.2422/2036-2145.201010_011}
  {\bibfield  {journal} {\bibinfo  {journal} {Ann. Sc. Norm. Super. Pisa Cl.
  Sci.}\ }\textbf {\bibinfo {volume} {12}},\ \bibinfo {pages} {209} (\bibinfo
  {year} {2013})}\BibitemShut {NoStop}%
\bibitem [{\citenamefont {Coleman}(1988)}]{Coleman:1987rm}%
  \BibitemOpen
  \bibfield  {author} {\bibinfo {author} {\bibfnamefont {S.~R.}\ \bibnamefont
  {Coleman}},\ }\bibfield  {title} {\bibinfo {title} {{Quantum Tunneling and
  Negative Eigenvalues}},\ }\href
  {https://doi.org/10.1016/0550-3213(88)90308-2} {\bibfield  {journal}
  {\bibinfo  {journal} {Nucl. Phys. B}\ }\textbf {\bibinfo {volume} {298}},\
  \bibinfo {pages} {178} (\bibinfo {year} {1988})}\BibitemShut {NoStop}%
\bibitem [{\citenamefont {Basar}\ \emph {et~al.}(2013)\citenamefont {Basar},
  \citenamefont {Dunne},\ and\ \citenamefont {Unsal}}]{Basar:2013eka}%
  \BibitemOpen
  \bibfield  {author} {\bibinfo {author} {\bibfnamefont {G.}~\bibnamefont
  {Basar}}, \bibinfo {author} {\bibfnamefont {G.~V.}\ \bibnamefont {Dunne}},\
  and\ \bibinfo {author} {\bibfnamefont {M.}~\bibnamefont {Unsal}},\ }\bibfield
   {title} {\bibinfo {title} {{Resurgence theory, ghost-instantons, and
  analytic continuation of path integrals}},\ }\href
  {https://doi.org/10.1007/JHEP10(2013)041} {\bibfield  {journal} {\bibinfo
  {journal} {JHEP}\ }\textbf {\bibinfo {volume} {10}},\ \bibinfo {pages}
  {041}},\ \Eprint {https://arxiv.org/abs/1308.1108} {arXiv:1308.1108 [hep-th]}
  \BibitemShut {NoStop}%
\bibitem [{\citenamefont {Witten}(2011)}]{Witten:2010cx}%
  \BibitemOpen
  \bibfield  {author} {\bibinfo {author} {\bibfnamefont {E.}~\bibnamefont
  {Witten}},\ }\bibfield  {title} {\bibinfo {title} {{Analytic continuation of
  Chern-Simons theory}},\ }\href@noop {} {\bibfield  {journal} {\bibinfo
  {journal} {AMS/IP Stud. Adv. Math.}\ }\textbf {\bibinfo {volume} {50}},\
  \bibinfo {pages} {347} (\bibinfo {year} {2011})},\ \Eprint
  {https://arxiv.org/abs/1001.2933} {arXiv:1001.2933 [hep-th]} \BibitemShut
  {NoStop}%
\bibitem [{\citenamefont {Tanizaki}\ and\ \citenamefont
  {Koike}(2014)}]{Tanizaki:2014xba}%
  \BibitemOpen
  \bibfield  {author} {\bibinfo {author} {\bibfnamefont {Y.}~\bibnamefont
  {Tanizaki}}\ and\ \bibinfo {author} {\bibfnamefont {T.}~\bibnamefont
  {Koike}},\ }\bibfield  {title} {\bibinfo {title} {{Real-time Feynman path
  integral with Picard-Lefschetz theory and its applications to quantum
  tunneling}},\ }\href {https://doi.org/10.1016/j.aop.2014.09.003} {\bibfield
  {journal} {\bibinfo  {journal} {Annals Phys.}\ }\textbf {\bibinfo {volume}
  {351}},\ \bibinfo {pages} {250} (\bibinfo {year} {2014})},\ \Eprint
  {https://arxiv.org/abs/1406.2386} {arXiv:1406.2386 [math-ph]} \BibitemShut
  {NoStop}%
\bibitem [{\citenamefont {Behtash}\ \emph {et~al.}(2017)\citenamefont
  {Behtash}, \citenamefont {Dunne}, \citenamefont {Sch{\"a}fer}, \citenamefont
  {Sulejmanpasic},\ and\ \citenamefont {{\"U}nsal}}]{Behtash:2015loa}%
  \BibitemOpen
  \bibfield  {author} {\bibinfo {author} {\bibfnamefont {A.}~\bibnamefont
  {Behtash}}, \bibinfo {author} {\bibfnamefont {G.~V.}\ \bibnamefont {Dunne}},
  \bibinfo {author} {\bibfnamefont {T.}~\bibnamefont {Sch{\"a}fer}}, \bibinfo
  {author} {\bibfnamefont {T.}~\bibnamefont {Sulejmanpasic}},\ and\ \bibinfo
  {author} {\bibfnamefont {M.}~\bibnamefont {{\"U}nsal}},\ }\bibfield  {title}
  {\bibinfo {title} {{Toward Picard-Lefschetz theory of path integrals, complex
  saddles and resurgence}},\ }\href
  {https://doi.org/10.4310/AMSA.2017.v2.n1.a3} {\bibfield  {journal} {\bibinfo
  {journal} {Ann. Math. Sci. Appl.}\ }\textbf {\bibinfo {volume} {2}},\
  \bibinfo {pages} {95} (\bibinfo {year} {2017})},\ \Eprint
  {https://arxiv.org/abs/1510.03435} {arXiv:1510.03435 [hep-th]} \BibitemShut
  {NoStop}%
\bibitem [{\citenamefont {Blum}\ and\ \citenamefont
  {Mirbabayi}(2024)}]{Blum:2024hcs}%
  \BibitemOpen
  \bibfield  {author} {\bibinfo {author} {\bibfnamefont {K.}~\bibnamefont
  {Blum}}\ and\ \bibinfo {author} {\bibfnamefont {M.}~\bibnamefont
  {Mirbabayi}},\ }\bibfield  {title} {\bibinfo {title} {{A single-bubble source
  for gravitational waves in a cosmological phase transition}},\ }\href
  {https://doi.org/10.1088/1475-7516/2024/08/039} {\bibfield  {journal}
  {\bibinfo  {journal} {JCAP}\ }\textbf {\bibinfo {volume} {08}},\ \bibinfo
  {pages} {039}},\ \Eprint {https://arxiv.org/abs/2403.20164} {arXiv:2403.20164
  [hep-ph]} \BibitemShut {NoStop}%
\bibitem [{\citenamefont {Liu}\ \emph {et~al.}(2025)\citenamefont {Liu},
  \citenamefont {Ma}, \citenamefont {Wei},\ and\ \citenamefont
  {Wu}}]{Liu:2025}%
  \BibitemOpen
  \bibfield  {author} {\bibinfo {author} {\bibfnamefont {Y.}~\bibnamefont
  {Liu}}, \bibinfo {author} {\bibfnamefont {X.-N.}\ \bibnamefont {Ma}},
  \bibinfo {author} {\bibfnamefont {J.}~\bibnamefont {Wei}},\ and\ \bibinfo
  {author} {\bibfnamefont {W.}~\bibnamefont {Wu}},\ }\bibfield  {title}
  {\bibinfo {title} {{Entire solutions of the magnetic Ginzburg--Landau
  equation in $\mathbb{R}^4$}},\ }\href
  {https://doi.org/10.2422/2036-2145.202209_005} {\bibfield  {journal}
  {\bibinfo  {journal} {Ann. Sc. Norm. Super. Pisa Cl. Sci.}\ }\textbf
  {\bibinfo {volume} {26}},\ \bibinfo {pages} {757} (\bibinfo {year}
  {2025})}\BibitemShut {NoStop}%
\bibitem [{\citenamefont {Gidas}\ \emph {et~al.}(1979)\citenamefont {Gidas},
  \citenamefont {Ni},\ and\ \citenamefont {Nirenberg}}]{Gidas:1979}%
  \BibitemOpen
  \bibfield  {author} {\bibinfo {author} {\bibfnamefont {B.}~\bibnamefont
  {Gidas}}, \bibinfo {author} {\bibfnamefont {W.-M.}\ \bibnamefont {Ni}},\ and\
  \bibinfo {author} {\bibfnamefont {L.}~\bibnamefont {Nirenberg}},\ }\bibfield
  {title} {\bibinfo {title} {{Symmetry and related properties via the maximum
  principle}},\ }\href {https://doi.org/10.1007/BF01221125} {\bibfield
  {journal} {\bibinfo  {journal} {Commun. Math. Phys.}\ }\textbf {\bibinfo
  {volume} {68}},\ \bibinfo {pages} {209} (\bibinfo {year} {1979})}\BibitemShut
  {NoStop}%
\bibitem [{\citenamefont {Zhang}\ \emph {et~al.}(2026)\citenamefont {Zhang},
  \citenamefont {Cui}, \citenamefont {Jiang}, \citenamefont {Shi},
  \citenamefont {Zhang}, \citenamefont {Yin},\ and\ \citenamefont
  {Zhang}}]{Zhang:2025}%
  \BibitemOpen
  \bibfield  {author} {\bibinfo {author} {\bibfnamefont {Z.}~\bibnamefont
  {Zhang}}, \bibinfo {author} {\bibfnamefont {G.}~\bibnamefont {Cui}}, \bibinfo
  {author} {\bibfnamefont {K.}~\bibnamefont {Jiang}}, \bibinfo {author}
  {\bibfnamefont {A.-C.}\ \bibnamefont {Shi}}, \bibinfo {author} {\bibfnamefont
  {P.}~\bibnamefont {Zhang}}, \bibinfo {author} {\bibfnamefont
  {J.}~\bibnamefont {Yin}},\ and\ \bibinfo {author} {\bibfnamefont
  {L.}~\bibnamefont {Zhang}},\ }\bibfield  {title} {\bibinfo {title}
  {{Exploring transition pathways in the Landau-Brazovskii model}},\ }\href
  {https://doi.org/10.1103/q4gq-dkkp} {\bibfield  {journal} {\bibinfo
  {journal} {Phys. Rev. Res.}\ }\textbf {\bibinfo {volume} {8}},\ \bibinfo
  {pages} {013196} (\bibinfo {year} {2026})},\ \Eprint
  {https://arxiv.org/abs/2510.17080} {arXiv:2510.17080} \BibitemShut {NoStop}%
\bibitem [{\citenamefont {Benfenati}\ \emph {et~al.}(2020)\citenamefont
  {Benfenati}, \citenamefont {Maiani}, \citenamefont {Rybakov},\ and\
  \citenamefont {Babaev}}]{Benfenati:2019}%
  \BibitemOpen
  \bibfield  {author} {\bibinfo {author} {\bibfnamefont {A.}~\bibnamefont
  {Benfenati}}, \bibinfo {author} {\bibfnamefont {A.}~\bibnamefont {Maiani}},
  \bibinfo {author} {\bibfnamefont {F.~N.}\ \bibnamefont {Rybakov}},\ and\
  \bibinfo {author} {\bibfnamefont {E.}~\bibnamefont {Babaev}},\ }\bibfield
  {title} {\bibinfo {title} {{Vortex nucleation barrier in superconductors
  beyond the Bean-Livingston approximation: A numerical approach for the
  sphaleron problem in a gauge theory}},\ }\href
  {https://doi.org/10.1103/PhysRevB.101.220505} {\bibfield  {journal} {\bibinfo
   {journal} {Phys. Rev. B}\ }\textbf {\bibinfo {volume} {101}},\ \bibinfo
  {pages} {220505} (\bibinfo {year} {2020})},\ \Eprint
  {https://arxiv.org/abs/1911.09513} {arXiv:1911.09513 [cond-mat.supr-con]}
  \BibitemShut {NoStop}%
\end{thebibliography}%

\clearpage
\onecolumngrid
\begin{center}
\textbf{Supplemental Material}
\end{center}
\twocolumngrid
\setcounter{figure}{0}
\renewcommand{\thefigure}{S\arabic{figure}}

\begin{figure*}[t!]
\includegraphics[width=0.95\columnwidth]{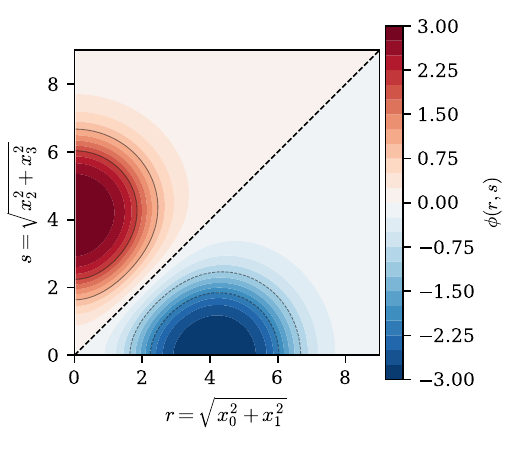}
\includegraphics[width=0.95\columnwidth]{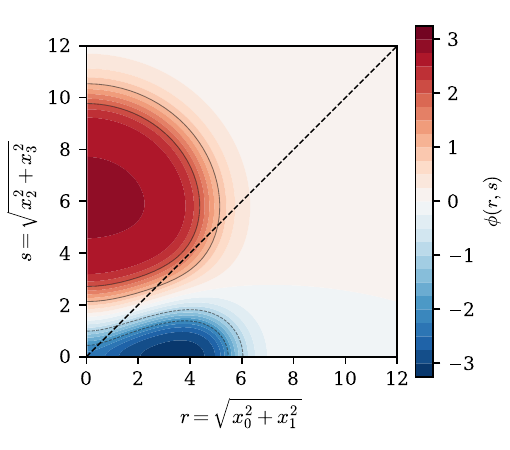}
\caption{Left: The nonradial solution for the potential of \eq{eq:V} at $g=0.1$, on the reduced half-plane $(r,s)$ with $r=\sqrt{x_0^2+x_1^2}$, $s=\sqrt{x_2^2+x_3^2}$. The field is odd under the exchange $r\leftrightarrow s$ (dashed line). Each lump is the cross section of a bubble tube wrapping a ring of radius $R_*\simeq4.15$ in $\mathbb{R}^4$: a positive-sign ring in the $(x_2,x_3)$ plane and its negative mirror image in the $(x_0,x_1)$ plane. Right: The nonradial solution for the parity-violating potential of \eq{eq:kappa_potential} at $\kappa=g=0.1$.}
\label{fig:solution}
\end{figure*}


\begin{figure}[t]
\includegraphics[width=0.95\columnwidth]{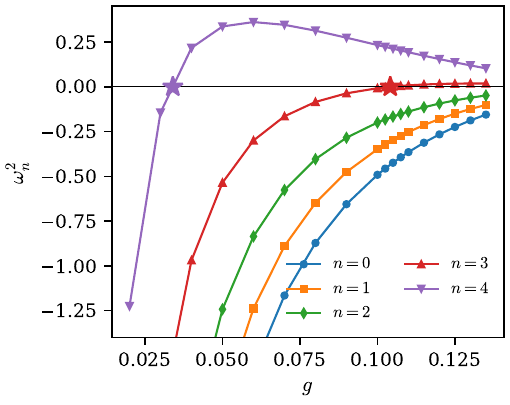}
\caption{Lowest fluctuation eigenvalue in the sectors $(0,n)$, $n=0,\dots,4$, along the branch. Each curve that crosses zero (stars) sheds a quadruplet of negative modes: the count of negative modes is 19 for $g\lesssim0.034$, 15 for $0.034\lesssim g\lesssim0.104$, and 11 for $g\gtrsim0.104$.
}
\label{fig:modes}
\end{figure}

\textit{Numerical methods.---}On the reduced $(r,s)$ half-plane the action reads
\begin{equation}
S=4\pi^2\!\int\! dr\,ds\; r s\left[\tfrac{1}{2}(\partial_r\phi)^2+\tfrac{1}{2}(\partial_s\phi)^2+V(\phi)\right],
\end{equation}
and the equation of motion is $\partial_r^2\phi+\frac{1}{r}\partial_r\phi+\partial_s^2\phi+\frac{1}{s}\partial_s\phi=V'(\phi)$. We discretize on a cell-centered grid, $r_i=(i+\frac12)h$, with a conservative second-order stencil and the measure $r_i s_j h^2$, in a box of size $L$ with $\phi=0$ on the outer boundary. The antisymmetry $\phi(r,s)=-\phi(s,r)$ is enforced exactly at every step. Newton steps solve the linearized problem with preconditioned GMRES: the two-dimensional Laplacian is a tensor sum of one-dimensional tridiagonal operators, so its exact inverse (after a similarity transformation that symmetrizes the radial operators) is available by dense eigendecomposition of the one-dimensional factors, and serves as the preconditioner. Stalled steps fall back to Tikhonov damping. Converged residuals reach the double-precision roundoff floor of the discretized operator, below $10^{-11}$ in the measure-weighted norm. At fixed $g$, every Newton seed we tried either relaxes to the trivial configuration $\phi=0$ or converges to the nonradial solution described in the main text, consistent with the class minimizer guaranteed by~\cite{Mederski:2017,Jeanjean:2018}. \fig{fig:solution} shows the solution on the $(r,s)$ plane.

Each $g$ is solved on two grids, $h=0.1$ and $h=0.067$, and the action is Richardson-extrapolated assuming $O(h^2)$ convergence; at $g=0.1$ the extrapolated action lies $0.6\%$ above the $h=0.1$ value, and the same procedure is consistent between the two grids at the $10^{-4}$ level. The box size, $L=20$ for most of the branch, is enlarged automatically when the solution leaks to the boundary (e.g.\ $L=28$ at $g=0.135$). The branch is continued in $g$ by seeding Newton with the neighboring solution, with an analytic fallback seed built from the O(3) bounce profile bent into rings of the extrapolated radius. Continuation fails for $g\gtrsim 0.14$: the thin-wall limit narrows the Newton basin, and seeds we tried either stall or relax to the trivial solution. 

Fluctuation spectra per $(m,n)$ sector are computed with a preconditioned block eigensolver (LOBPCG) on the same grids, using the same tensor-eigenbasis preconditioner, with the centrifugal terms $m^2/r^2+n^2/s^2$ included in the one-dimensional factors. The lower end of the spectrum is shown in \fig{fig:modes}. The translation and rotation zero modes come out at the discretization floor ($\lesssim5\times10^{-4}$ at $h=0.1$, scaling as $h^2$), and we verified the rotational quadruplet against the analytic form $r\partial_s\phi-s\partial_r\phi$ underlying all broken angular momentum generators. The $n=3$ negative eigenvalue, which has the smallest absolute magnitude at $g=0.1$, is stable under box enlargement ($L=20\to26$ changes it by less than $10^{-5}$) and under grid refinement (Richardson value $-0.0078$). Eigenvalues quoted along the branch for $n=3,4$ are Richardson-extrapolated from the two grids; the remaining sectors are quoted at $h=0.1$.

The straight-tube collapse eigenvalue $\omega_0^2$ entering~\eq{eq:dispersion} is the lowest radial eigenvalue of the fluctuation operator around the O(3) bounce, computed by shooting plus a one-dimensional eigensolve. The O(4) and O(3) bounces themselves are obtained by shooting with bisection on the core value, with the asymptotic tail $\phi\propto e^{-\rho}\rho^{-(d-1)/2}$ attached analytically; the O(4) action satisfies the virial identity to six digits.

\begin{figure*}[t]
\includegraphics[width=0.98\textwidth]{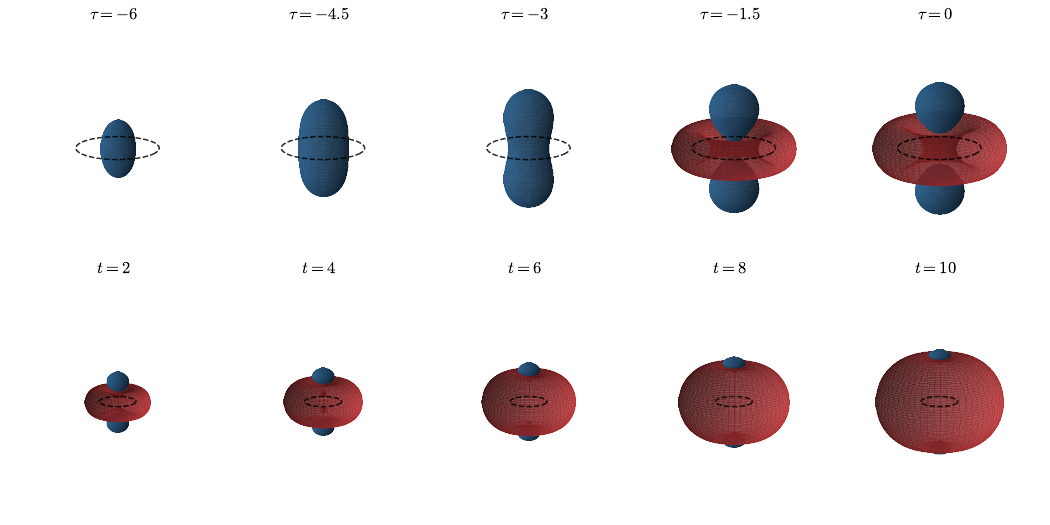}
\caption{Constant-field surfaces $|\phi|=1$ of the saddle at $g=0.1$ (red: $\phi=+1$; blue: $\phi=-1$). Top row: Euclidean development, on slices $x_0=\tau$. The slice first grazes the negative tube, which lies in the $(x_0,x_1)$ plane (a single blob at $\tau=-6$); the blob stretches and splits into two anti-sign bubbles while the positive torus pulls out of the vacuum, reaching the time-symmetric configuration at $\tau=0$. Bottom row: Lorentzian evolution of the $\tau=0$ Cauchy data. The dashed circle, drawn identically in every panel, is the initial ring core, of radius $R_*\simeq4.15$ in the $(x_2,x_3)$ plane; it sets the common physical scale (the bottom-row panels span a region about twice as large as the top row). The positive ring expands and fills in, approaching an asymptotically spherical true-vacuum bubble; the anti-sign bubbles expand more slowly as dimples near the poles of the expanding wall. The discrete leapfrog energy is conserved at the $0.2\%$ level of the gross energy budget through $t=10$.}
\label{fig:evolution}
\end{figure*}

\textit{Euclidean development and real-time evolution.---}The solution is even in each Cartesian coordinate, so the slice $x_0=0$ carries time-symmetric Cauchy data $(\phi,\dot\phi=0)$ for the Lorentzian field equation. The Euclidean-time energy $E_E(\tau)=\int d^3x\,[-\frac{1}{2}(\partial_\tau\phi)^2+\frac{1}{2}(\nabla\phi)^2+V]$ is conserved along $x_0$ and vanishes as $x_0\to-\infty$ where $\phi\to0$; at $\tau=0$, where $\partial_\tau\phi=0$, it equals the Lorentzian energy of the slice data, which therefore vanishes exactly. On the lattice we find the gradient and potential integrals of the slice data equal and opposite ($\pm848.0$ at $g=0.1$) to four digits. Like Coleman's bounce, the configuration is degenerate with the false vacuum and materializes at rest.

Fig.~\ref{fig:evolution} shows the Euclidean development of the configuration (top row) and its real-time evolution (bottom row). On the slice the field is axisymmetric about the $x_1$ axis, $\phi=\phi(t;x_1,\sqrt{x_2^2+x_3^2})$, and the Lorentzian evolution preserves this symmetry, reducing to a $2{+}1$-dimensional wave equation that we evolve with a leapfrog scheme on the Euclidean grid ($h=0.067$, $dt=0.03$, conservative cylindrical stencil). The discrete energy, evaluated in summation-by-parts form at the staggered half-steps, starts at $-0.04$ (consistent with zero up to discretization) and is conserved to $0.2\%$ of the gross kinetic-plus-gradient budget through $t=10$; late-time accuracy is limited by Lorentz contraction of the walls. The qualitative outcome is that the positive ring behaves as a super-critical bubble: it expands, its hole fills in, and the wall approaches a sphere at late times, while the two anti-sign bubbles riding on its axis expand more slowly and are progressively engulfed. The evolution also exhibits the residual Lorentz symmetry of the solution: the Euclidean O(2) acting on the $(x_0,x_1)$ plane continues to SO(1,1), so the anti-sign bubbles, which descend from the ring $x_0^2+x_1^2=R_*^2$, must trace the boost-invariant hyperbola $x_1^2-t^2=R_*^2$. The measured core positions follow $x_1(t)=\sqrt{R_*^2+t^2}$ to $0.1\%$ throughout the run---a nontrivial check.

\textit{Negative-mode flows and the thimble question.---}Whether a Euclidean saddle contributes to a given path integral is, in Picard--Lefschetz language, the question of whether its upward-flow cycle intersects the original integration cycle~\cite{Witten:2010cx,Tanizaki:2014xba,Behtash:2015loa}. 
We computed these flows at $g=0.1$: $\partial_\eta\phi=\nabla^2\phi-V'(\phi)$, with $\eta$ the flow parameter, seeded at $\phi_{\rm saddle}\pm\epsilon\,\psi$ for each negative mode $\psi$ of the $(0,0)$ sector ($\epsilon=0.15$, $|\psi|_{\max}=1$), evolved with an IMEX scheme on the same grid (Laplacian implicit through the tensor solver, $d\eta=0.1$). The $P$-odd dilation mode ($\omega^2=-0.315$) flows to the false vacuum for one sign of $\epsilon$ and to unbounded true-vacuum growth ($S\to-\infty$) for the other: the saddle is a transition state between the false vacuum and the escape region, with the same connection topology as the O(4) bounce's single negative mode. The two $P$-even modes ($\omega^2=-0.492,\,-0.090$) flow to escape for both signs; the two signs are exact mirror images under $P$ combined with $\phi\to-\phi$ (their action histories agree to machine precision), so these directions connect escape to escape. The bending quadruplets $(0,n)$ break one O(2) and would require three-dimensional flows, which we have not computed.

Because the negative-mode count is odd on the whole branch, the Gaussian factor $i^k$ is imaginary and a nonvanishing intersection number contributes to ${\rm Im}\,E$ with relative weight $e^{-(\Sodd-\SOfour)}$. The flow topology above supports a nonzero index; a controlled evaluation, including the resolution of the Stokes ambiguity, is left open.

\end{document}